\documentclass[preprint,preprintnumbers, prd, floatfix, superscriptaddress,nofootinbib] {revtex4-1}
\usepackage{graphicx}
\usepackage{epsfig}
\usepackage{bm}
\usepackage{amssymb}
\usepackage{float}
\usepackage{amsmath}
\usepackage{dcolumn}
\usepackage{cancel}
\usepackage[usenames,dvipsnames]{color}
\usepackage{color}
\usepackage{epstopdf}
\usepackage{nccbbb}

\def\doi{http://doi.org}

\begin{document}

\title{Constraints on running vacuum model with $H(z)$ and $f \sigma_8$}

\author{Chao-Qiang Geng}
\email{geng@phys.nthu.edu.tw}
\affiliation{Chongqing University of Posts \& Telecommunications, Chongqing, 400065, China
}
\affiliation{Department of Physics, National Tsing Hua University, Hsinchu, Taiwan 300}
\affiliation{National Center for Theoretical Sciences, Hsinchu,
Taiwan 300}

\author{Chung-Chi Lee}
\email{lee.chungchi16@gmail.com}
\affiliation{
DAMTP, Centre for Mathematical Sciences, University of Cambridge, Wilberforce Road, Cambridge CB3 0WA, UK}

\author{Lu Yin}
\email{yinlumail@foxmail.com}
\affiliation{Department of Physics, National Tsing Hua University, Hsinchu, Taiwan 300}

\begin{abstract}
We examine  the running vacuum model with $\Lambda (H) = 3 \nu H^2 + \Lambda_0$, where $\nu$ is the model parameter
and $\Lambda_0$ is the cosmological constant. From the data of the cosmic microwave background radiation, weak lensing and
baryon acoustic oscillation along with the time dependent Hubble parameter $H(z)$ and weighted linear growth $f (z)\sigma_8(z)$ measurements, we find that $\nu=(1.37^{+0.72}_{-0.95})\times 10^{-4}$ with the best fitted $\chi^2$ value slightly smaller than that in the $\Lambda$CDM model.
%In addition, we show that the allowed windows for the neutrino mass sum $\Sigma m_{\nu}$ and  the mass fluctuation amplitude related to $\sigma_8$ are suppressed .
\end{abstract}

\maketitle

\section{Introduction}

To understand the accelerated expansion of the universe discovered in 1998~\cite{Riess:1998cb, Perlmutter:1998np}, dark energy is introduced ~\cite{Copeland:2006wr}.
In a variety scenarios of dark energy, the Lambda cold dark matter ($\Lambda$CDM) model is the simplest one, which can explain the
 current cosmological observations very well. Unfortunately, this model accompanies several theoretical difficulties, such as ``fine-tuning''~\cite{Weinberg:1988cp, WBook} and ``coincidence''~\cite{Ostriker:1995rn, ArkaniHamed:2000tc} problems.

The running vacuum model (RVM) is one of the popular attempts to solve the ``coincidence''  problem~\cite{Ozer:1985ws, Carvalho:1991ut, Lima:1994gi, Lima:1995ea, Overduin:1998zv, Dymnikova:2001ga, Carneiro:2004iz, Bauer:2005rpa, Alcaniz:2005dg, Barrow:2006hia, Shapiro:2009dh, Geng:2016epb, Geng:2016dqe}.
In this model, the cosmological constant $\Lambda$ is described by a function of the Hubble parameter and decays to matter and radiation in the expansion of the universe~\cite{Ozer:1985ws}.
It has been shown that the RVM is suitable in describing the cosmological evolution on both background and linear perturbation levels~\cite{Shapiro:2004ch, Arcuri:1993pb, EspanaBonet:2003vk, Basilakos:2009wi, Sola:2013gha, Costa:2012xw, Gomez-Valent:2014rxa, Sola:2014tta, Tamayo:2015qla, Fritzsch:2016ewd}.

In this work, we focus on the specific model with the running cosmological constant
$\Lambda = 3\nu H^2 + \Lambda_0$, where $\nu$ is the model parameter
and $\Lambda_0$ is the cosmological constant in the $\Lambda$CDM model. Clearly, in this RVM, the $\Lambda$CDM limit corresponds to $\nu=0$.  Naively, it is expected that the value of $\nu$ should  be arbitrarily close to zero in order to fit the current cosmological observations. However, it has been recently shown that $\nu \sim \mathcal{O}(10^{-2})- \mathcal{O}(10^{-3})$
in this RVM with the exclusion of
the $\Lambda$CDM model within $2 \sigma$ confidence level in the literature~\cite{Sola:2016vis, Sola:2016jky, Gomez-Valent:2014fda, Sola:2015wwa, Gomez-Valent:2015pia, Sola:2016ecz},
which indicates that this model is a better theory to describe the evolution history of our universe.
In this study, we plan to reexamine this RVM by using the latest observational data.
In particular, we include the measurements of  the time dependent Hubble parameter $H(z)$ and weighted linear growth $f (z)\sigma_8(z)$ in our analysis. We will use the {\bf CAMB}~\cite{Lewis:1999bs} and {\bf CosmoMC}~\cite{Lewis:2002ah} packages with the Markov chain Monte Carlo (MCMC) method.

This paper is organized as follows.
In Sec.~\ref{sec:model}, we introduce the RVM and derive the evolution equations for matter and radiation in the linear perturbation theory.
In Sec.~\ref{sec:observation}, we perform the numerical calculations to obtain the observational constraints
on the model parameter $\nu$ and  cosmological observables in several  datasets.
Finally, our conclusions are given in Sec.~\ref{sec:CONCLUSIONS}.

\section{THE RUNNING VACUUM MODEL}
\label{sec:model}

\subsection{Formalism}
We start from the Einstein equation,
\begin{equation}
\label{eq:action}
R_{\mu\nu}-\frac{1}{2}Rg_{\mu\nu}+\Lambda g_{\mu\nu} = 8\pi G T^{M}_{\mu\nu},
\end{equation}
where $R=g^{\mu\nu}R_{\mu\nu}$,  $\Lambda$ and $T^{M}_{\mu\nu}$  are the Ricci scalar,
 cosmological constant and energy-momentum tensor of matter and radiation, respectively.
Considering the Friedmann-Lemaitre-Robertson-Walker (FLRW) metric
\begin{eqnarray}
ds^{2}= -dt^{2}+ a^{2}(t) \delta_{ij}dx^{i}dx^{j} \,,
\end{eqnarray}
we obtain the Friedmann equations,
\begin{eqnarray}
\label{eq:friedmann1}
&& H^{2}=\frac{8\pi G}{3}(\rho_m + \rho_r +\rho_{\Lambda}) \,, \\
\label{eq:friedmann2}
&& \dot{H}=- 4\pi G(\rho_m + \rho_r +\rho_{\Lambda} + P_m+ P_r + P_{\Lambda}), \,
\end{eqnarray}
where $H=da/(adt)$ is the Hubble parameter, $\rho_\alpha$ ($P_\alpha$)
%, $\rho_r$ ($P_r$) and $\rho_{\Lambda}$ ($P_{\Lambda}$)
with $\alpha=r,m$ and $\Lambda$
represent the energy densities (pressures) of matter, radiation and dark energy, respectively.
Furthermore, it is convenient to define the corresponding equations of state, given by
\begin{eqnarray}
\label{eq:eos}
w_{m,r,\Lambda}=\frac{P_{m,r,\Lambda}}{\rho_{m,r,\Lambda}}= 0, \frac{1}{3}, -1 \,.
\end{eqnarray}

In the RVM, dark energy decays to radiation and matter in the evolution of the universe,
so that the continuity equations can be written as,
\begin{eqnarray}
\label{eq:continuity}
&& \dot{\rho}_M+3 H(1+w_M)\rho_M = Q \,, \\
&& \dot{\rho}_\Lambda+3 H(1+w_\Lambda)\rho_\Lambda = - Q \,,
\end{eqnarray}
where $\rho_M = \rho_m + \rho_r$, $w_M = (P_m + P_r) / \rho_M$ and $Q = Q_m+Q_r$ with $Q_{m(r)}$ the decay rate of dark energy to matter (radiation).
In this work, we consider $\Lambda$ to be a function of the Hubble parameter, which might originate from the cosmological renormalization group~\cite{Sola:2013gha}, given by
\begin{equation}
\label{eq:rhode}
\Lambda = 3 \nu H^2+\Lambda_0,
\end{equation}
where $\nu$ and $\Lambda_{0}$ are two free parameters
with the $\Lambda$CDM model  recovered by taking $\nu=0$.
If dark energy only couples to matter (radiation), there will be too many non-relativistic (relativistic) particles created in the early (late) time of the universe in terms of the current observations.
By combining Eqs.~\eqref{eq:continuity} and \eqref{eq:rhode}, the coupling $Q_\alpha$ ($\alpha$ = $m$ or $r)$ is given by
\begin{eqnarray}
\label{eq:coupling}
Q_\alpha=-\frac{\dot{\rho}_{\Lambda}(\rho_\alpha+P_\alpha)}{\rho_{M}+P_M}=3\nu H(1+w_\alpha)\rho_\alpha \,,
\end{eqnarray}
with $P_M = P_m + P_r$, where $\alpha$ represents matter or radiation.
As a result, the energy density of $\alpha$ can be derived from Eq.~\eqref{eq:continuity} as,
\begin{eqnarray}
\rho_\alpha= \rho^{(0)}_\alpha a^{-3(1+w_\alpha)\xi} \,,
\end{eqnarray}
where $\xi=1-\nu$ and $\rho^{(0)}_\alpha$ is the energy density at $z=0$.
Note that $\nu \geq 0$ is chosen to avoid the negative dark energy density in the early universe.

\subsection{Linear perturbation theory}
\label{sec:P}
We follow the standard linear perturbation theory~\cite{Ma:1995ey} and derive the growth equation of the density perturbation in the RVM.
The metric with the synchronous gauge is given by,
\begin{eqnarray}
ds^{2}=a^{2}(\tau)[-d\tau^{2}+(\delta_{ij}+h_{ij})dx^{i}dx^{j}],
\end{eqnarray}
with  $i,j=1,2,3$, $\tau$  the conformal time and
\begin{eqnarray}
h_{ij}= \int d^3 k e^{i\vec k\vec x}[\hat{k}_i \hat{k}_j h(\vec k,\tau)+6(\hat{k}_i\hat{k}_j-\frac{1}{3}\delta_{ij})\eta(\vec k,\tau)]                   ,
\end{eqnarray}
where $h(\vec k,\tau)$ and $\eta(\vec k,\tau)$ are two scalar perturbations, and $\hat{k}= \vec k/k$ is the $k$-space unit vector.
By using the conservation equation $\nabla^{\nu}( T^{M}_{\mu\nu}+T^{\Lambda}_{\mu\nu})=0$ with $\delta T^{0}_{0}=\delta \rho_{M}$, $\delta T^{0}_{i}=-T^{i}_{0}=(\rho_{M}+P_{M})v^i_M$ and $\delta T^{i}_{j}=\delta P_{M}\delta ^{i}_{j}$,
one gets the growth equations of the matter and radiation as follows,
\begin{eqnarray}
\label{eq:pert1}
&& \dot{\delta_\alpha}=-(1+w_{\alpha})(\theta_{\alpha}+\frac{\dot{h}}{2})-3H(\frac{\delta P_{\alpha}}{\delta \rho_{\alpha}}-w_{\alpha})\delta_{\alpha}-\frac{Q_{\alpha}}{\rho_{\alpha}}\delta_{\alpha} \,, \\
\label{eq:pert2}
&& \dot{\theta}_{\alpha}=-H(1-3w_{\alpha})\theta_{\alpha}-\frac{\dot{w}_{\alpha}}{1+w_{\alpha}}\theta_{\alpha}+\frac{\delta P_{\alpha}/\delta \rho_{\alpha}}{1+w_{\alpha}}\frac{k^{2}}{a^{2}}\delta_{\alpha}-\frac{Q_{\alpha}}{\rho_{\alpha}}\theta_{\alpha} \,,
\end{eqnarray}
where $\delta_\alpha\equiv \delta \rho_\alpha/\rho_\alpha$ and $\theta_\alpha=ik_{i}v^{i}_\alpha$ are the density fluctuation and the divergence of fluid velocity, respectively.

%We note that the dark energy perturbation has been discussed in Refs.~\cite{Fabris:2006gt, Borges:2008ii, VelasquezToribio:2009qp, Geng:2016fql, Gomez-Valent:2014rxa, Sola:2015rra, Sola:2015wwa, Gomez-Valent:2015pia, Basilakos:2015vra, Sola:2016jky, Sola:2016ecz, Sola:2016zeg, Sola:2017jbl}, in which the Hubble parameter is rewritten to be a Lorentz scalar with $H = \nabla_{\mu} U^{\mu}/3$ and $U^{\mu}=dx^{\mu}/ds$.
%However, the expression of $H$ is not unique, and the cosmological behavior significantly depends on the explicit form.
%In order not to lose the generality, we concentrate on the  homogeneous and isotropic dark energy universe.
In principle, the dark energy density fluctuation should be taken into account when the dynamical model is considered.
To explore the dynamics of dark energy, the time dependent cosmological constant should be rewritten as a Lorentz scalar at the field equation level in Eq.~\eqref{eq:action}.
For example, in Refs.~\cite{Fabris:2006gt, Borges:2008ii, VelasquezToribio:2009qp, Geng:2016fql, Gomez-Valent:2014rxa, Sola:2015rra, Sola:2015wwa, Gomez-Valent:2015pia, Basilakos:2015vra, Sola:2016jky, Sola:2016ecz, Sola:2016zeg, Sola:2017jbl}, the vacuum energy 
is given
 as $\Lambda = \Lambda(H)$ with $H=\nabla_{\mu} U^{\mu}/3$ and $U^{\mu}=dx^{\mu}/ds$.
However, the dark energy perturbation plays an important role to the evolution of $\delta_m$ at the subhorizon scale, leading to a strong interaction between  dark energy and  matter fields, which must be ruled out by the astrophysical observations.
In addition, the expression of $H$ is not unique, and the cosmological behavior significantly depends on the explicit form.
Due to these two reasons and without losing the generality, we concentrate on the homogeneous and isotropic dark energy model, i.e., $\delta \rho_{\Lambda} = 0$.
Consequently, we have $\delta_{\Lambda} = \theta_\Lambda = 0$, so that the particles, created from the dark energy decays, homogeneously distribute to the universe, smoothing the density fluctuation by the factor $Q_\alpha / \rho_\alpha$.

In the RVM, due to the background evolution of the Hubble parameter, one has
\begin{eqnarray}
\label{eq:Hz}
\frac{H^2}{H_0^2} = \frac{\Omega_m a^{-3 \xi} + \Omega_r a^{-4 \xi} + \Omega_\Lambda^\star}{1-\nu} \,,
\end{eqnarray}
where $\Omega_{m(r)} = \rho_{m(r)}^{(0)} / 3H_0^2$, $\Omega_\Lambda^\star = \Omega_\Lambda - \nu = \rho_\Lambda^{(z=0)} / 3H_0^2 - \nu $ and $\Omega_m + \Omega_r + \Omega_\Lambda =1$.
As discussed in Ref.~\cite{Fritzsch:2016ewd}, the larger $\nu$ is, the smaller $H(z)$ behaves in the high redshift regime.

It is known that the spectrum of the cosmic matter fluctuations can give important constraints on theoretical models
about the structure formation
~\cite{Press:1973iz,Bond:1990iw,Bower:1991kf,Kauffmann:1993jn,Lacey:1993iv}.
These fluctuations can be described by the weighted linear growth $f(z)\sigma_8 (z)$, where
\begin{eqnarray}
f(z)&=&-(1+z){d\ln\delta_{m}\over dz}
\end{eqnarray}
 is the growth factor and $\sigma_8 (z)$ is the root-mean-square matter fluctuation amplitude on the scale
 of $R_8= 8h^{-1}$ Mpc at the redshift $z$, given by
\begin{eqnarray}
\sigma^{2}_8 (z)= \delta^2_{m}(z)\int \frac{d^3 k}{(2\pi)^3}P(k,\vec p)W^2(kR_8)\,,
\end{eqnarray}
with
$P(k,\vec p)$  the ordinary linear matter power spectrum and
$W(kR_8)$  the top-hat smoothing function (see e.g.\cite{Gomez-Valent:2014rxa} for details).
Several methods have been used to estimate $\sigma_8$, such as the measurements of the abundance of galaxy clusters~\cite{Eke:1996ds,Borgani:2001ir,Reiprich:2001zv,Pierpaoli:2000ip}, cosmic shear analyses~\cite{Bacon:2002va,Brown:2002wt},  combined analysis of galaxy redshift survey~\cite{Bond:2002tp} and CMB data~\cite{Lahav:2001sg}.

We note that the RVM modifies not only the background evolution but also the perturbation one.
The creations of matter and radiation from the decays of dark energy suppress the growths of the density fluctuations, as demonstrated in Eqs.~\eqref{eq:pert1} and \eqref{eq:pert2}.
If $\nu$ is large, the suppression effect on $\delta_m$ should be significant, leading to a ``lowering effect'' on
$f(z) \sigma_8(z)$~\cite{Gomez-Valent:2014rxa, Sola:2015rra, Sola:2015wwa, Gomez-Valent:2015pia, Basilakos:2015vra, Sola:2016jky, Sola:2016ecz, Sola:2016zeg, Sola:2017jbl}.
Clearly, it is interesting to examine the RVM by using the data from the large scale structure observations, such as the baryon acoustic oscillation (BAO) and $f \sigma_8$.

\section{Numerical calculations}
\label{sec:observation}

\begin{table}[!hbp]
\caption{$H(z)$ data points}
\begin{tabular}{|c|c|c|c||c|c|c|c||c|c|c|c|}
\hline
\ & $z$ & $H(z)$  & Ref. & \ & $z$ & $H(z)$   & Ref. & \ & $z$ & $H(z)$ 
  & Ref. \\
 \ &  &   ($km/s/Mpc$) &  & \ &  &   ($km/s/ Mpc)$ & & \ & & 
 ($km/s/Mpc$) &  \\
\hline
~1~ & 0.07 & 69.0$\pm$19.6 & ~\cite{Zhang:2012mp} &
~13~ & 0.4 & 95.0$\pm$17.0 &~\cite{Simon:2004tf} &
~25~ & 0.9 & 117.0$\pm$23.0 & ~\cite{Simon:2004tf} \\
\hline
2 & 0.09 & 69.0$\pm$12.0 & ~\cite{Jimenez:2003iv} & 14 & 0.4004 & 77.0$\pm$10.2 &~\cite{Moresco:2016mzx} &26 & 1.037 & 154.0$\pm$20.0 & ~\cite{Moresco:2012jh}\\
\hline
3 & 0.12 & 68.6$\pm$26.2 & ~\cite{Zhang:2012mp} &15 & 0.4247 & 87.1$\pm$11.2 & ~\cite{Moresco:2016mzx}&27 & 1.3 & 168.0$\pm$17.0 & ~\cite{Simon:2004tf}\\
\hline
4 & 0.17 & 83.0$\pm$8.0 & ~\cite{Simon:2004tf} &16 & 0.4497 & 92.8$\pm$12.9 & ~\cite{Moresco:2016mzx}& 28 & 1.363 & 160.0$\pm$33.6 & ~\cite{Moresco:2015cya}\\
\hline
5 & 0.179 & 75.0$\pm$4.0 & ~\cite{Moresco:2012jh} &17 & 0.4783 & 80.9$\pm$9.0 & ~\cite{Moresco:2016mzx}& 29 & 1.43 & 177.0$\pm$18.0 &  ~\cite{Simon:2004tf}\\
\hline
6 & 0.199 & 75.0$\pm$5.0 & ~\cite{Moresco:2012jh} &18 & 0.48 & 97.0$\pm$62.0 &~\cite{Stern:2009ep} & 30 & 1.53 & 140.0$\pm$14.0 &  ~\cite{Simon:2004tf}\\
\hline
7 & 0.2 & 72.9$\pm$29.6 & ~\cite{Zhang:2012mp} &19 & 0.57 & 92.4$\pm$4.5 & ~\cite{Reid:2012sw}&31 & 1.75 & 202.0$\pm$40.0 &  ~\cite{Simon:2004tf} \\
\hline
8 & 0.27 & 77.0$\pm$14.0 & ~\cite{Simon:2004tf} &20 & 0.5929 & 104.0$\pm$13.0 & ~\cite{Moresco:2012jh} & 32 & 1.965 & 186.5$\pm$50.4 & ~\cite{Moresco:2015cya}\\
\hline
9 & 0.24 & 79.69$\pm$2.65 &  ~\cite{Gaztanaga:2008xz} &21 & 0.6797 & 92.0$\pm$8.0 & ~\cite{Moresco:2012jh}& 33 & 2.3 & 224$\pm$8 & ~\cite{Busca:2012bu}\\
\hline
10 & 0.28 & 88.8$\pm$36.6 & ~\cite{Zhang:2012mp} &22 & 0.7812 & 105.0$\pm$12.0 & ~\cite{Moresco:2012jh}&34 & 2.34 & 222$\pm$7 & ~\cite{Hu:2014vua}\\
\hline
11 & 0.352 & 83.0$\pm$14.0 & ~\cite{Moresco:2012jh} & 23 & 0.8754 & 125.0$\pm$17.0 & ~\cite{Moresco:2012jh}& 35 & 2.36 & 226$\pm$8 &~\cite{Font-Ribera:2013wce} \\
\hline
~12~ & 0.3802 & 83.0$\pm$13.5 & ~\cite{Moresco:2016mzx} &24 & 0.88 & 90.0$\pm$40.0 &~\cite{Stern:2009ep}& & & & \\
\hline
\end{tabular}
\label{tab:1}
\end{table}

\begin{table}[!hbp]
\caption{$f\sigma_8$ data points}
\begin{tabular}{|c|c|c|c||c|c|c|c||c|c|c|c|}
\hline
\ & $z$ & $f\sigma_8$ & Ref. & \ & $z$ & $f\sigma_8$ & Ref. & \ & $z$ & $f\sigma_8$ & Ref.  \\
\hline
~$1$~ & $1.36$ & $0.482 \pm 0.116$ & \cite{Okada:2015vfa} &
~$10$~ & $0.59$ & $0.488 \pm 0.06$ & \cite{Chuang:2013wga} &
~$19$~ & $0.35$ & $0.440 \pm 0.05$ & \cite{Song:2008qt, Tegmark:2006az}\\
\hline
$2$ & $0.8$ & $0.470 \pm 0.08$ & \cite{delaTorre:2013rpa} &
$11$ & $0.57$ & $0.444 \pm 0.038$ & \cite{Gil-Marin:2015sqa} &
$20$ & $0.32$ & $0.394 \pm 0.062$ & \cite{Gil-Marin:2015sqa} \\
\hline
$3$ & $0.78$ & $0.38 \pm 0.04$ & \cite{Blake:2011rj} &
$12$ & $0.51$ & $0.452 \pm 0.057$ & \cite{Satpathy:2016tct} &
$21$ & $0.3$ & $0.407 \pm 0.055$ & \cite{Tojeiro:2012rp} \\
\hline
$4$ & $0.77$ & $0.490 \pm 0.18$ & \cite{Song:2008qt, Guzzo:2008ac} &
$13$ & $0.5$ & $0.427 \pm 0.043$ & \cite{Tojeiro:2012rp} &
$22$ & $0.25$ & $0.351 \pm 0.058$ & \cite{Samushia:2011cs}\\
\hline
$5$ & $0.73$ & $0.437 \pm 0.072$ & \cite{Blake:2012pj} &
$14$ & $0.44$ & $0.413 \pm 0.080$ & \cite{Blake:2012pj}&
$23$ & $0.22$ & $0.42 \pm 0.07$ & \cite{Blake:2011rj}\\
\hline
$6$ & $0.61$ & $0.457 \pm 0.052$ & \cite{Satpathy:2016tct} &
$15$ & $0.41$ & $0.45 \pm 0.04$ & \cite{Blake:2011rj} &
$24$ & $0.17$ & $0.51 \pm 0.06$ & \cite{Song:2008qt, Percival:2004fs} \\
\hline
$7$ & $0.60$ & $0.390 \pm 0.063$ & \cite{Blake:2012pj} &
$16$ & $0.4$ & $0.419 \pm 0.041$ & \cite{Tojeiro:2012rp} &
$25$ & $0.15$ & $0.49 \pm 0.15$ & \cite{Howlett:2014opa} \\
\hline
$8$ & $0.6$ & $0.433 \pm 0.067$ & \cite{Tojeiro:2012rp} &
$17$ & $0.38$  & $0.430 \pm 0.054$ & \cite{Satpathy:2016tct} &
$26$ & $0.067$ & $0.423 \pm 0.055$ & \cite{Beutler:2012px} \\
\hline
$9$ & $0.60$ & $0.43 \pm 0.04$ & \cite{Blake:2011rj} &
$18$ & $0.37$ & $0.460 \pm 0.038$ & \cite{Samushia:2011cs} &
$27$ & $0.02$ & $0.36 \pm 0.04$ & \cite{Hudson:2012gt} \\
\hline
\end{tabular}
\label{tab:2}
\end{table}

In Tables~\ref{tab:1} and \ref{tab:2}, we list 35 and 27 points for $H(z)$ and $f(z)\sigma_8(z)$ from
 the time dependent  Hubble parameter and large scale structure formation measurements, respectively.
By performing the {\bf CosmoMC} program~\cite{Lewis:2002ah},
we fit the RVM from the observational data with the MCMC method.
The dataset includes those from $H(z)$ and $f(z) \sigma_8(z)$ along with the
CMB temperature fluctuation from {\it Planck 2015} with TT, TE, EE, low-$l$
polarization from SMICA~\cite{Adam:2015wua, Aghanim:2015xee, Ade:2015zua}, the weak lensing (WL) data from CFHTLenS ~\cite{Heymans:2013fya} and the  BAO data from 6dF Galaxy Survey~\cite{Beutler:2011hx} and BOSS~\cite{Anderson:2013zyy}.
In addition, the $\chi^2$ function for the data from $H(z)$ or $f(z) \sigma_8(z)$ is taken to be
\begin{eqnarray}
\chi^2_c = \sum_{i=1}^n \frac{(T_c(z_i) - O_c(z_i))^2}{E^i_c} \,,
\end{eqnarray}
where the subscript $c$, representing $H(z)$ or $f(z) \sigma_8(z)$, 
denotes the category of the data at the redshift $z_i$ from different data, $n$ is the number of the data in each dataset, $T_c$ is the theoretical prediction, calculated from {\bf CAMB}, and $O_c$ ($E_c$) is the observational value (covariance).
The priors of the various parameters are listed in Table~\ref{tab:3}.

\begin{table}[ht]
\begin{center}
\caption{ Priors for cosmological parameters with $\Lambda= 3\nu H^2 + \Lambda_0$.  }
\begin{tabular}{|c||c|} \hline
Parameter & Prior
\\ \hline
Model parameter $\nu$& $0 \leq \nu \leq 3 \times 10^{-4}$
\\ \hline
Baryon density & $0.5 \leq 100\Omega_bh^2 \leq 10$
\\ \hline
CDM density & $10^{-3} \leq \Omega_ch^2 \leq 0.99$
\\ \hline
Optical depth & $0.01 \leq \tau \leq 0.8$
\\ \hline
Neutrino mass sum& $0 \leq \Sigma m_{\nu} \leq 2$~eV
\\ \hline
$\frac{\mathrm{Sound \ horizon}}{\mathrm{Angular \ diameter \ distance}}$  & $0.5 \leq 100 \theta_{MC} \leq 10$
\\ \hline
Scalar power spectrum amplitude & $2 \leq \ln \left( 10^{10} A_s \right) \leq 4$
\\ \hline
Spectral index & $0.8 \leq n_s \leq 1.2$
\\ \hline
\end{tabular}
%\vskip 0.2in
\label{tab:3}
\end{center}
\end{table}

\begin{figure}
\centering
\includegraphics[width=0.96 \linewidth]{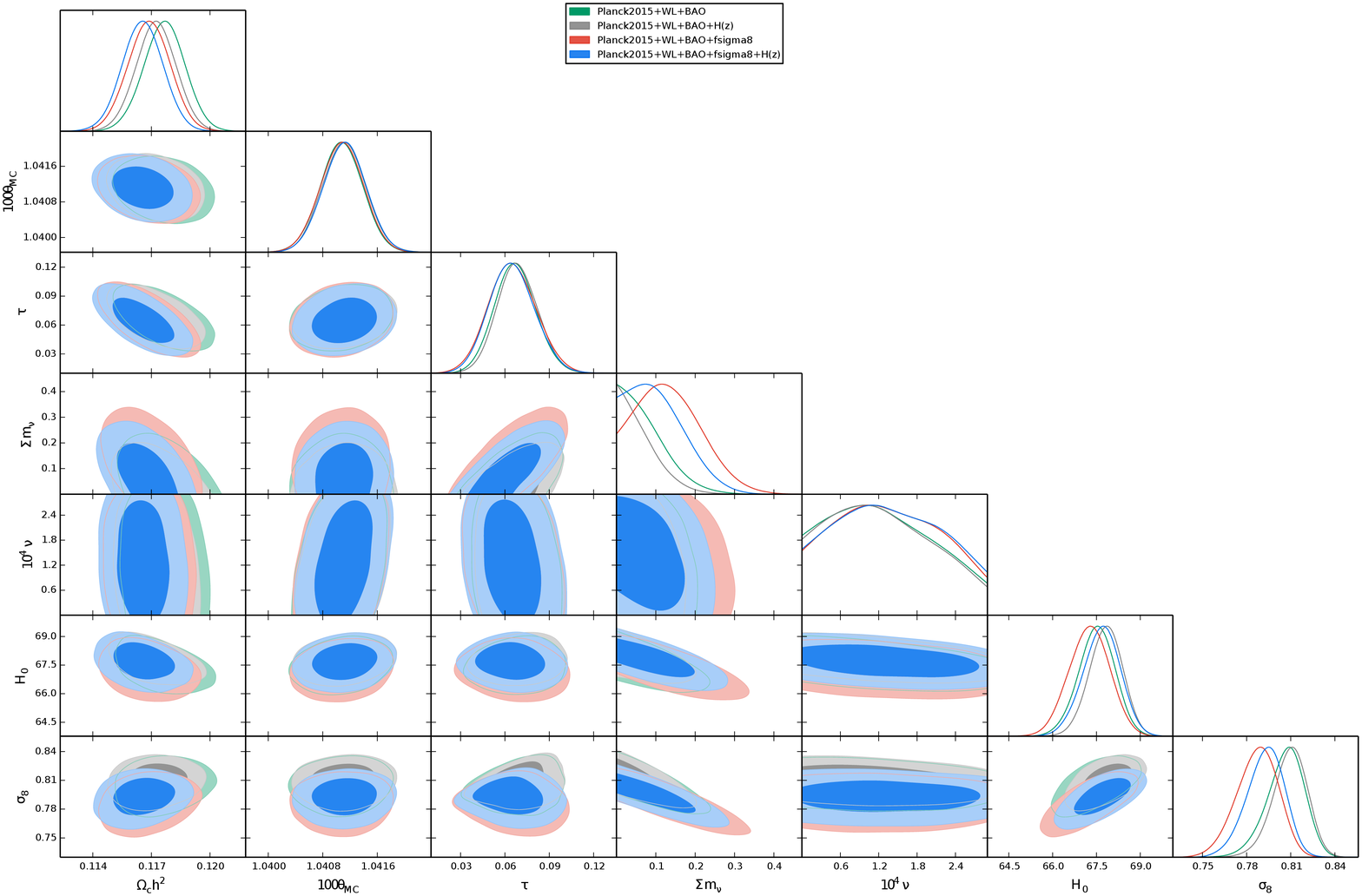}
\caption{One and two-dimensional distributions of $\Omega_c h^2$, $100\theta_{MC}$, $\tau$, $\sum m_\nu$, $10^4\nu$, $\sigma_8$, where the contour lines represent 68$\%$~ and 95$\%$~ C.L., respectively.}
\label{fg:2}
\end{figure}

In Fig.~\ref{fg:2} and Table~\ref{tab:4}, we present our global fit result from various datasets, and the values in the brackets correspond to the best-fit results in the $\Lambda$CDM model, where the values of $\sigma_8$  are given at $z=0$.
%In particular, from the combined data of the CMB, WL, BAO, $H(z)$ and $f (z)\sigma_8(z)$, we find that $\nu=(1.37^{+0.72}_{-0.95})\times 10^{-4}$ with the best fitted $\chi^2$ value being 13531.2, which is slightly smaller than 13534.7 in the $\Lambda$CDM model.
From the combined datasets of  (A), (B), (C) and (D), we find that $\nu =1.35$, 1.91,1.55 and $1.23\times 10^{-4}$ with the best fitted $\chi^2$ values being 13487.7, 13509.9, 13511.3 and 13531.2 in the RVM,  which are all slightly smaller than 13488.9, 13512.2, 13512.8 and 13534.7 in the $\Lambda$CDM model, respectively.
This combined dataset leads to the lowest $\chi^2$ value in comparison with that in the $\Lambda$CDM model as shown in Table~\ref{tab:4}.
We note that a part  of the $H(z)$ data in Table~\ref{tab:1} is derived by using the
BC03 stellar models~\cite{Bruzual:2003tq}, which  shows tensions from those by the MaStro models~\cite{Maraston:2011sq} at high redshift regime, as pointed out in Ref.~\cite{Moresco:2012jh}.
Specifically, the $H(z)$ data point at $z = 1.037$ in the MaStro models is $1.6\sigma$ lower than that in the BC03 one.
From Eq.~\eqref{eq:Hz}, we can estimate that with the same initial conditions $H(z)$ with $\nu_1$ is smaller than that with $\nu_2$ if $\nu_1 > \nu_2$, so that the constraint on $\nu$ should be slightly relaxed when the BC03 data are replaced by the MaStro ones.
\begin{table}[!hbp]
\newcommand{\tabincell}[2]{\begin{tabular}{@{}#1@{}}#2\end{tabular}}
\caption{The $1D$ marginalized fitting results for the RVM with $\Lambda$ = 3$\nu H^2$+$\Lambda_0$, where the limits are given at 95$\%$ C.L. with $\nu$  calculated within 68$\%$ C.L., and the numbers in the bracket represent the central values in the $\Lambda$CDM model.}
\begin{tabular}{|c|c|c|c|c|}
\hline
Parameter &
\tabincell{c}{ (A) \\{\it Planck} + \\  WL + BAO}  &
\tabincell{c}{ (B) \\ {\it Planck} + \\  WL + BAO + $f\sigma_8$} &
\tabincell{c}{ (C) \\{\it Planck} + \\  WL + BAO + $H(z)$} &
\tabincell{c}{ (D) \\{\it Planck} + WL+ \\  BAO + $f\sigma_8$ + $H(z)$} \\
\hline
%\tabincell{c}{ Model parameter \\ $10^4\nu$ } & $1.29^{+0.54}_{-1.12}$ & $1.36^{+0.73}_{-0.96}$ & $1.27^{+0.53}_{-1.10}$ & $1.37^{+0.72}_{-0.95}$ \\
\tabincell{c} { Model parameter \\ $10^4\nu$ } & $<1.83$ & $<2.09$ & $<1.80$ & $<2.09$ \\
\hline
\tabincell{c}{ Baryon density \\ $100 \Omega_bh^2$ } & $ 2.23 \pm 0.03$ (2.23) & $2.23^{+0.04}_{-0.03}$ (2.24) & $2.23^{+0.02}_{-0.03} $ (2.23) & $ 2.22 \pm 0.03$ (2.24) \\
\hline
\tabincell{c}{ CDM density \\ $100 \Omega_ch^2$ } & $11.8 \pm 0.2$ (11.8) & $11.7 \pm 0.2$ (11.7) & $11.7 \pm 0.2$ (11.7) & $11.7^{+0.2}_{-0.3}$ (11.7) \\
\hline
\tabincell{c}{ Optical depth \\ $100 \tau$ } & $6.67^{+2.83}_{-2.70}$ (6.96) & $6.48^{+3.23}_{-3.03}$ (6.99) & $6.84^{+2.76}_{-2.61}$ (7.13) & $6.49^{+3.08}_{-2.91}$ (6.96) \\
\hline
$\sigma_8$ & $0.806^{+0.025}_{-0.026}$ (0.810) & $0.787^{+0.027}_{-0.028}$ (0.788) & $0.809^{+0.023}_{-0.024}$ (0.812) & $0.792^{+0.025}_{-0.026}$ (0.793) \\
\hline
\tabincell{c}{ Neutrino mass \\ $\Sigma m_\nu / \mathrm{eV}$ } & $< 0.188$ ($< 0.198$) & $< 0.278$ ($< 0.301$) & $< 0.161$($<0.176$) & $ <0.235 $ ($<0.262$) \\
\hline
$\chi^2_{best-fit}$ & $13487.7$ (13488.9) & $ 13509.9 $ (13512.2) & $ 13511.3 $ (13512.8) & $ 13531.2 $ (13534.7) \\
\hline
\end{tabular}
\label{tab:4}
\end{table}

Although the cosmological observables in the RVM do not significantly deviate from those in $\Lambda$CDM, the best $\chi^2$ fits in the RVM are better than those in the $\Lambda$CDM model in all the datasets.
It clearly indicates that the RVM is  good in describing the evolution history of our universe. %favored by the observations.
It should be noted that our result of  $\nu\sim \mathcal{O}(10^{-4})$ is about one to two orders stronger that those of
$\nu \sim \mathcal{O}(10^{-3})- \mathcal{O}(10^{-2})$ in the literature~\cite{Sola:2016vis, Sola:2016jky, Gomez-Valent:2014fda, Sola:2015wwa, Gomez-Valent:2015pia, Sola:2016ecz}.
Because the constraints on $\nu$ only slightly change when the $f(z) \sigma_8(z)$ and $H(z)$ data are taken into account, the RVM might be strongly constrained by the CMB temperature fluctuation.
Moreover, we are unable to exclude the $\Lambda$CDM model more than $1.5 \sigma$ confidence level, which is different
from the $2 \sigma$ exclusion statement in Refs.~\cite{Sola:2016vis, Sola:2016jky, Gomez-Valent:2014fda, Sola:2015wwa, Gomez-Valent:2015pia, Sola:2016ecz}.
%\textcolor{blue}{Optical depth can therefore be thought of as the opacity of a medium}.
Since the creation of particles from the decaying dark energy restrains the growths of $\delta_m$ %and $\sigma_8$
is also suppressed in the RVM.
In addition, it is known that the free streaming massive neutrinos suppress the matter density fluctuation, which also smoothen the density fluctuation.
As a result, as shown in Table.~\ref{tab:4}, the allowed window for $\Sigma m_{\nu}$ is further restricted.

\section{Conclusions}
\label{sec:CONCLUSIONS}

We have explored the allowed window for the model parameter in the RVM with $\Lambda(H) = 3\nu H^2 + \Lambda_0$  and shown
 that the constraint on the RVM becomes much very stronger with $\nu \sim  \mathcal{O}( 10^{-4})$ after considering the CMB temperature fluctuation along with the $H(z)$ and  $f (z)\sigma_8(z)$ measurements, which is different from the results in the previous studies in the literature,
which might be due to the accurate measurement on the photon power spectrum. Clearly,
 this statement should be further studied in the future.
Explicitly, the value of $\nu=(1.37^{+0.72}_{-0.95})\times 10^{-4}$ is found at 68\% C.L. by fitting the combined data of the CMB, WL, BAO, $H(z)$ and $f (z)\sigma_8(z)$.
We have also found that $\chi^2_{RVM} \lesssim \chi^2_{\Lambda \mathrm{CDM}}$ in all the datasets of
our discussions, denoting that the RVM is a good theory to describe the evolution of the universe at both background and linear perturbation levels.
However, concerning the existence of an extra degree of freedom in the RVM to  $\Lambda$CDM, it is hard to claim that the RVM
 is preferred by the observational data at the current stage yet.
In addition, since dark energy decays to matter and radiation in the evolution  of the universe, the matter density fluctuation $\delta_m$ is suppressed, leading to the best fitted value of $\Sigma m_{\nu}$ is relatively smaller than the corresponding one in the $\Lambda$CDM model.

\section*{Acknowledgments}

We thank Prof. J. Sola for useful comments and discussions.
This work was partially supported by National Center for Theoretical Sciences, MoST (MoST-104-2112-M-007-003-MY3 and MoST-106-2917-I-564-055) and the Newton International Fellowship (NF160058) from the Royal Society (UK).

\end{document}